# Recent advances of defect-induced spin and valley polarized states in graphene

Yu Zhang (张钰)[1,2,*], Liangguang Jia (贾亮广)[1], Yaoyao Chen (陈瑶瑶)[1], Lin He (何林)[3], and Yeliang Wang (王业亮)[1,*]

[1]School of Integrated Circuits and Electronics, MIIT Key Laboratory for Low-Dimensional Quantum Structure and Devices, Beijing Institute of Technology, Beijing 100081, People's Republic of China.

[2]Advanced Research Institute of Multidisciplinary Sciences, Beijing Institute of Technology, Beijing 100081, People's Republic of China.

[3]Center for Advanced Quantum Studies, Department of Physics, Beijing Normal University, Beijing 100875, People's Republic of China.

[*]Corresponding author. E-mail: yzhang@bit.edu.cn

[*]Corresponding author. E-mail: yeliang.wang@bit.edu.cn

**Electrons in graphene have fourfold spin and valley degeneracies owing to the unique bipartite honeycomb lattice and an extremely weak spin-orbit coupling, which can support a series of broken symmetry states. Atomic-scale defects in graphene are expected to lift these degenerate degrees of freedom at the nanoscale, and hence, lead to rich quantum states, highlighting promising directions for spintronics and valleytronics. In this article, we mainly review the recent scanning tunneling microscopy (STM) advances on the spin and/or valley polarized states induced by an individual atomic-scale defect in graphene, including a single-carbon vacancy, a nitrogen-atom dopant, and a hydrogen-atom chemisorption. Lastly, we give a perspective in this field.**

# 1. Introduction

Besides real spin, electrons in pristine graphene have multiple degrees of freedom including sublattice pseudospin and valley pseudospin, owing to the unique bipartite honeycomb lattice and an extremely weak spin-orbit coupling of graphene.[1-3] In the past decade, both experimental results and theoretical calculations revealed that an individual atomic-scale defect in graphene can generate various broken symmetry states at the nanoscale, and therefore, is expected to result in many novel quantum states.[4,5] For example, almost all the types of the individual atomic-scale defect, on the one hand, can be regarded as an atomic-scale scatter and is able to introduce intervalley scattering process, on the other hand, can locally break the sublattice symmetry of graphene, thus generating valley polarized states in the vicinity of the defects.[6] An individual hydrogen-atom chemisorbed on graphene leads to an out-of-plane atomic configuration, which significantly enhances the atomic spin-orbit coupling (SOC) of graphene, thus resulting in a fully spin and valley polarized states at the nanoscale.[7,8] Moreover, both an individual hydrogen-atom chemisorbed on graphene and an isolated single-carbon vacancy in graphene can generate local magnetic moments, and such magnetic moments can be controllably manipulated by tuning the gate voltage and the local curvature.[9-12] Therefore, atomic-scale defects in graphene provide ideal model systems to systemically study the novel quantum states, highlighting promising directions for spintronics and valleytronics.

In this article, we mainly review the recent scanning tunneling microscopy (STM) advances on the various types of individual atomic-scale defects in graphene, as demonstrated in Fig. 1(a). The defect-induced broken symmetry states, including spin and/or valley polarized states, can be directly captured by the scanning tunneling spectroscopy (STS) measurements (Fig. 1(b-e)). We first concentrate on the local magnetic moments induced by an individual single-carbon vacancy in graphene and hydrogen-atom chemisorbed on graphene. And then, we introduce the intervalley scattering process induced by an atomic-scale defect in monolayer and bilayer graphene, and point out the Berry phase signatures from the intervalley scattering patterns. After that, we introduce various spin and/or valley polarized states in graphene Landau level spectra induced by different types of atomic-scale defects. At last, we present the outlook and challenges in this field.

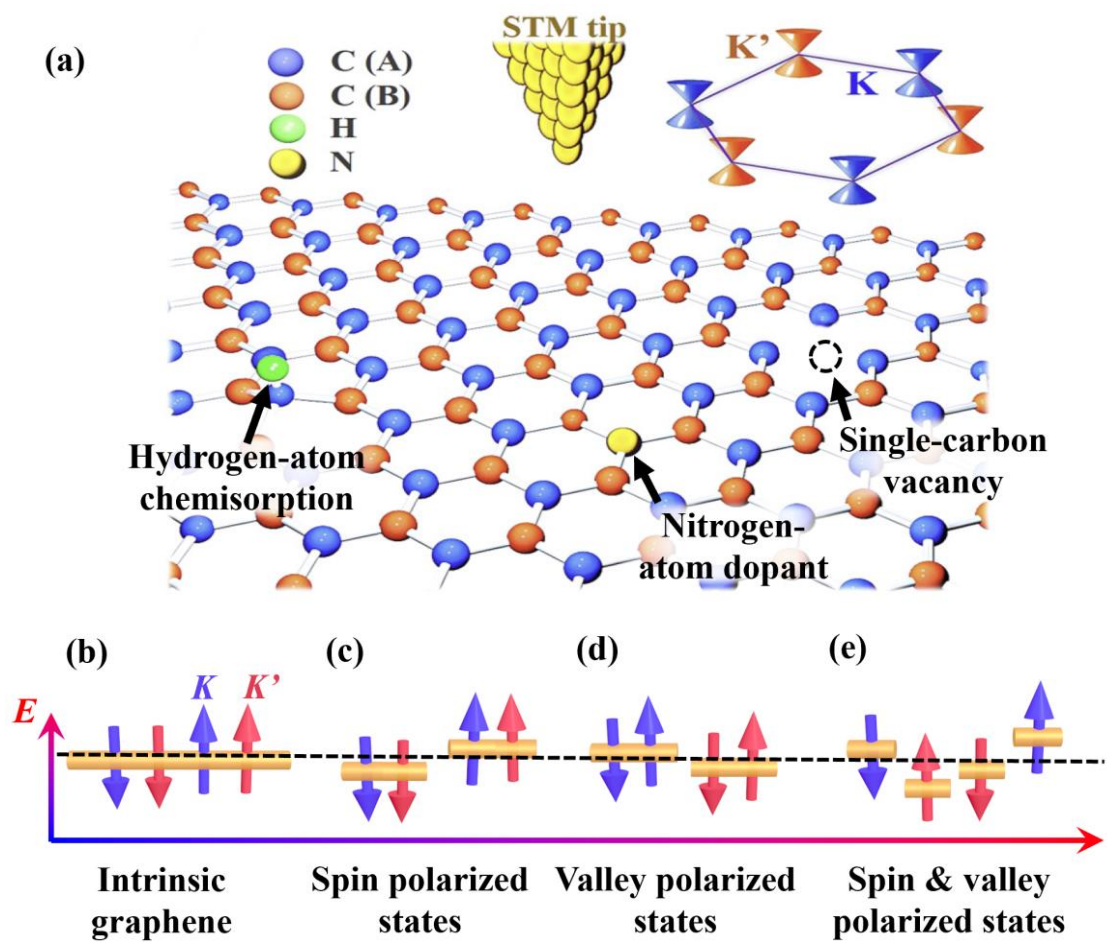

**Fig. 1.** Scanning tunneling microscopy (STM) study of the broken symmetry states induced by various of atomic-scale defects in graphene. (a) The studied atomic-scale defects include a single-carbon vacancy, a nitrogen-atom dopant, and a hydrogen-atom chemisorbed on graphene. (b) For the pristine graphene, there are fourfold spin and valley degeneracies. (c-e) Energy distributions of spin polarized, valley polarized, and spin & valley polarized states in graphene.

## 2. Local magnetic moment in graphene

According to Lieb's theorem, the ground state of materials with a bipartite lattice can support a magnetic moment when the numbers of sublattice sites are different.[13] Graphene has a unique bipartite honeycomb atomic structures consisting of two equivalent triangular A and B sublattices. Therefore, removing a single $p_z$ orbital from the graphene π electronic systems is expected to realize a local magnetic moment. In 2007, first principles calculations predicted that an individual hydrogen-atom chemisorbed on graphene can create a single π-state at the Fermi energy, which is expected to possess a local π magnetic moment of 1 $\mu_B$.[14] In contrast, an isolated single-carbon vacancy in graphene has a local magnetic moment of about 1.5 $\mu_B$, in which 1 $\mu_B$ is attributed to the unsaturated σ states and about 0.5 $\mu_B$ is attributed to the π electrons (Fig. 2).[14]

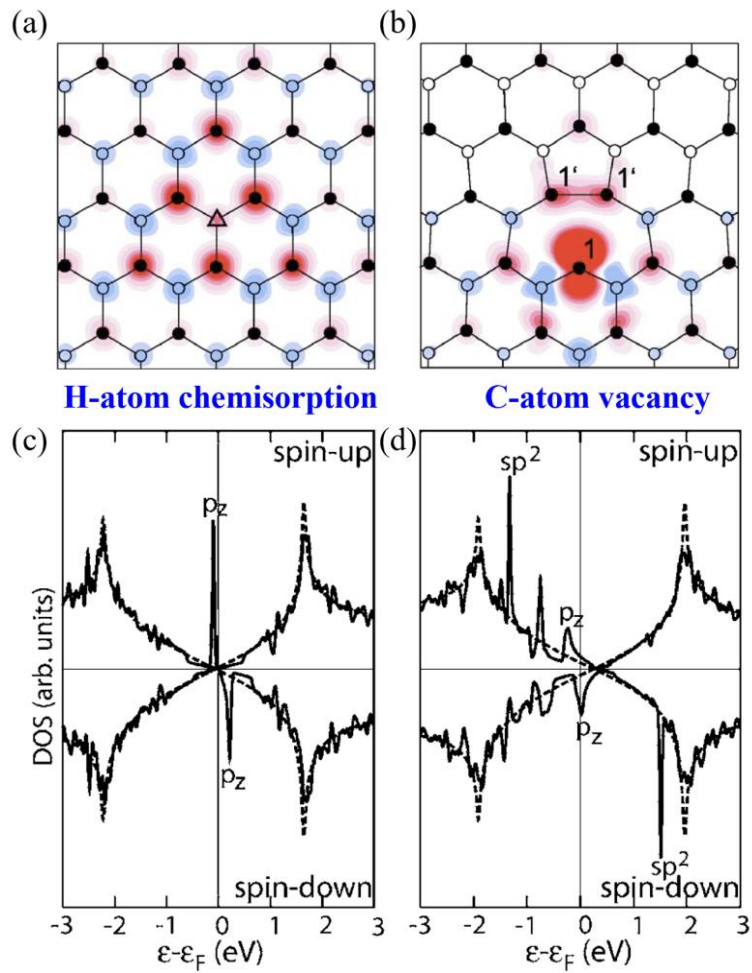

**Fig. 2.** First principles calculations of the atomic structures and the density of states around an individual single-carbon vacancy in graphene and hydrogen-atom chemisorbed on graphene. (a,b) Atomic structures around an individual single-carbon vacancy in graphene and hydrogen-atom chemisorbed on graphene. (c,d) Spin-density projection on the graphene around the an individual hydrogen-atom chemisorption and single-carbon vacancy. Reproduced with permission from Ref. [14]. Copyright 2007, American Physical Society.

Since the prediction, many experimental attempts have been adopted to study the local magnetic moments in graphene. In 2016, González-Herrero et al. provided the first experimental evidence of the local magnetic moment around an individual hydrogen-atom chemisorbed on graphene. The atomically resolved STM image shown in Fig. 3(a) exhibits a slight protrusion of the hydrogenated carbon atom and extremely small displacement of the surrounding carbon atoms. We can see from the scanning tunneling spectroscopy (STS) spectrum that the localized state induced by an individual hydrogen-atom chemisorption splits into two spin-polarized states, which contribute to the local magnetic moment, as depicted in Fig. 3(b).[9] It's worth noting that such spin-polarized states appear only when the localized state lying at the Fermi energy, and they vanish when the graphene system is either n-doping or p-doping (Fig. 3(c)). As a result, the local magnetic moment induced by an individual hydrogen-atom chemisorption is attributed to the electrostatic Coulomb repulsion, i.e., for the double occupation of the localized state by two electrons with different spins, once an electron occupies a state, the other one with opposite spin needs to pay an extra energy $U$, thus leading to a net magnetic moment, as demonstrated in Fig. 3(d).[15,16]

The atomic structures and magnetic states around the single-carbon vacancy in graphene are more complicated. Structurally, removing a single carbon atom in graphene creates three dangling σ bonds. Two of them experiences reconstruction and form a new σ bond, resulting in the local threefold symmetry broken down due to the in-plane Jahn-Teller distortion. The third dangling σ bond is left unsaturated and contributes to a local magnetic moment of 1 $\mu_B$.

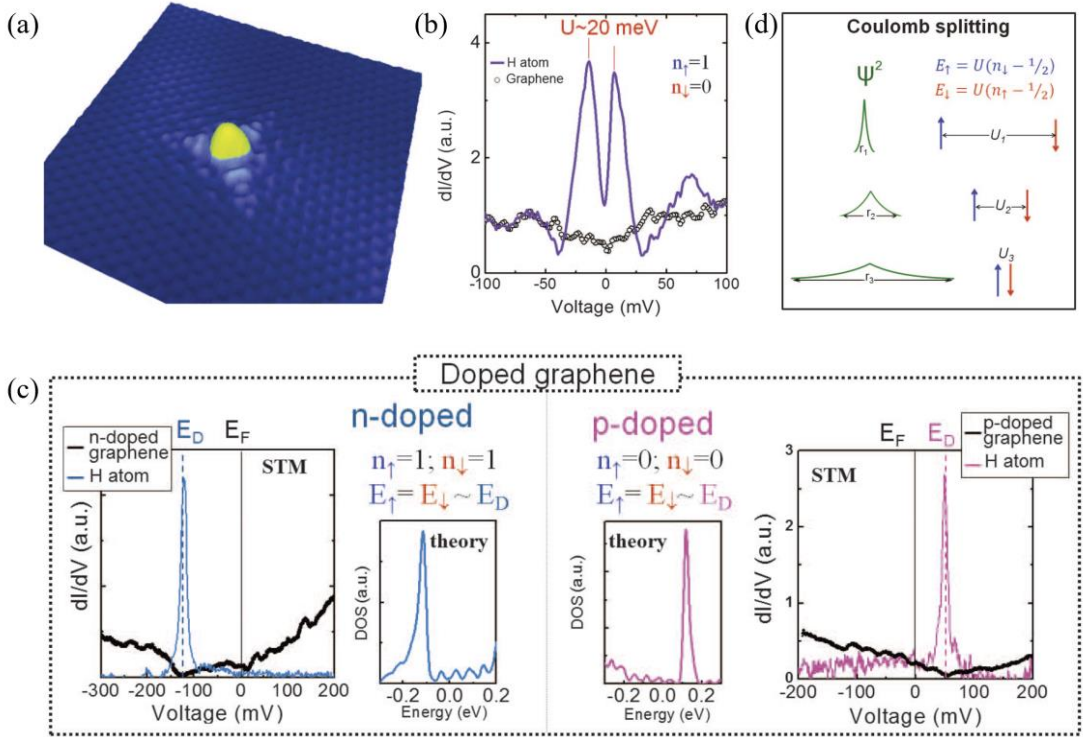

**Fig. 3.** Detection and manipulation of a local magnetic moment around an individual hydrogen-atom chemisorbed on graphene. (a) Atomic resolution STM image of a hydrogen-atom chemisorbed on graphene. (b) Representative STS spectrum recorded at the hydrogen-atom chemisorbed site. The two peaks reflect the density-of-states (DOS) with opposite spin polarizations. (c) Scanning tunneling spectroscopy (STS) spectra and density functional theory (DFT) calculation of the DOS induced by an individual hydrogen-atom chemisorbed on n-doped and p-doped graphene. (d) Illustration of the origin of the spin-split state around the hydrogen-atom chemisorption. Reproduced with permission from Ref. [9]. Copyright 2016, American Association for the Advancement of Science.

Experimentally, there are usually two approaches to create single-carbon vacancies in graphene. One approach is to irradiate graphene surface with high-energy ions, since the formation energy of a single-carbon vacancy in graphene is as high as about 7.4 eV.[17,18] In such a case, the single-carbon vacancies are usually located at the metastable state, instead of the ground state.[19-21] As a result, the measured STS spectra only exhibit one resonant peak at the Fermi energy,[19,22,23] and the measured average magnetic moment is only about 0.1 $\mu_B$ per vacancy,[24] much smaller than the expected value of about 1.5 $\mu_B$.[14] The other approach is to directly synthesize graphene on Ni/Rh foils with single-carbon vacancies via a facile ambient pressure chemical vapor deposition (CVD) method. During the fast-cooling process, a high density of single-carbon vacancies appear in the topmost graphene layer, owing to the segregation mechanism for graphene grown on Ni/Rh foils and the high carbon solubility of these metals.[10,11] Figure 4(a) shows a representative STM image of an isolated single-carbon vacancy in graphene, which exhibits a characteristic Jahn-Teller distortion around the vacancy. The STS spectrum recorded at the vacancy exhibits two pronounced peaks with their energy separation of about 30 meV, which are attributed to the two spin-polarized states of the π electrons (Fig. 4(b)).[10] Such spin-polarized states are quite robust and still exist in both n-doping and p-doping graphene, quite inconsistent with that in an individual hydrogen-atom chemisorption of graphene.

The interaction between the localized σ states and the quasi-localized π states determines the local magnetic moment generated by the single-carbon vacancy in graphene.[20,25,26] With the enhancement of the out-of-plane configuration around the single-carbon vacancy via the tip-graphene van der Waals (vdWs) force, the σ-π interactions can be efficiently tuned. As a consequence, the energy separations of two spin-polarized states decrease from about 30 meV down to about 20 meV, and finally reach 0 meV when the height of the protrusion increases to above 140 pm, as shown in Fig. 4(c).[11] These three magnetic states can be classified as ferromagnetic (FM), quenched antiferromagnetic (QAFM), and nonmagnetic (NM) phases. Specifically, for both the FM and QAFM phases, the localized σ electrons and the quasi-localized π electrons generate magnetic moments of about 1.0 $\mu_B$ and 0.6 $\mu_B$, respectively. However, the directions of the σ and π magnetic moments are parallel for the FM phase, yielding the total magnetic moments of about 1.6 $\mu_B$. In contrast, the directions of the σ and π magnetic moments are antiparallel for the QAFM phase, yielding the total magnetic moments of only about 0.5 $\mu_B$. In the NM phase, however, neither the σ nor π electrons exhibits local magnetic moment. Moreover, the coupling of local magnetic moment and the electrons in the conduction band can further drive a quantum phase transition from the local-moment phase to the Kondo-screened phase by using a gate voltage and a local curvature,[12,27] as shown in Fig. 4(d).

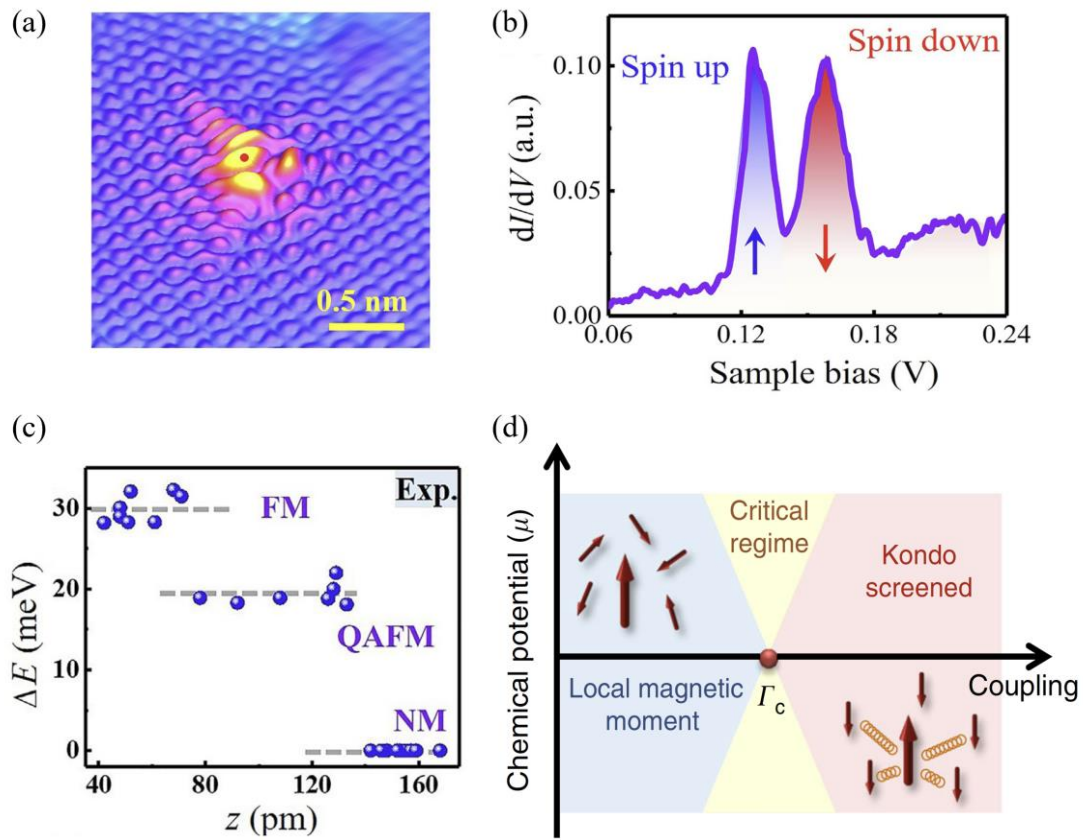

**Fig. 4.** Detection and manipulation of a local magnetic moment around an individual single-carbon vacancy in graphene. (a) Atomically resolved STM image of a single-carbon vacancy in graphene. (b) Representative STS spectrum recorded at the single-carbon vacancy in graphene. (c) The energy separations of the two spin-polarized DOS peaks in STS spectra as a function of the height of the vacancy in STM images, yielding three types of magnetic states, i.e., ferromagnetic (FM), quenched antiferromagnetic (QAFM), and nonmagnetic (NM) states. Reproduced with permission from Ref. [11]. Copyright 2020, Science China Press. (d) Phase transition from a local-magnetic-moment

phase to a Kondo-screened phase around a single-carbon vacancy in graphene. Reproduced with permission from Ref. [12]. Copyright 2018, Nature Publishing Group.

## 3. Local Berry phase signatures in graphene

Berry phase plays an important role in determining the topological properties of materials.[28-31] In the low-energy band structures of few-layer graphene systems, the nonzero Berry phase is associated with the pseudospin winding. For example, in monolayer graphene, the pseudospin rotates by the angle of $2\pi$ along a closed Fermi surface, which yields the winding number $W = 1$ and the Berry phase $\gamma = W\pi = \pi$, as shown in Fig. 5(a).[32] Previously, it is usually thought that measuring the Berry phase of materials requires the application of external magnetic fields, because the magnetic fields can force the electrons moving along a closed trajectory. Contradicting this belief, in 2019, Dutreix *et al.* took advantage of the intervalley scattering induced by an individual atomic-scale defect and realized to measure the Berry phase of monolayer graphene in the absence of external magnetic fields. In their framework, an atomic-scale defect in graphene can be regarded as an atomic scatterer that locally induces both intravalley and intervalley scattering processes.[33-40] Intravalley scattering process involves a π-rotation of the pseudospin, resulting in the interference destructive. In contrast, intervalley scattering process rotates the pseudospin by the angle of $2\theta_q$, where $\theta_q$ is the polar angle of electrons with the momentum $q$, as marked in Fig. 5(b). Therefore, the information about the pseudospin winding and the Berry phase can be directly captured from the intervalley scattering patterns.[38]

Figure 5(c) shows the real space representation of the intervalley scattering induced by an individual atomic-scale defect in monolayer graphene. At a given STM tip position, the electronic waves incident from the tip can be elastically scattered by the defect, with the pseudospin rotating by the angle of $2\theta_q$. In such a case, an atomic-scale defect in monolayer graphene can be regarded as a phase singularity, and further generates a pseudospin-mediated atomic-scale vortex with the angular momentum $|l| = 2$. Considering that the STM tip orientation relative to the defect can be expressed as $\theta_r = \theta_q + \pi$, the locking of $\theta_r$ and $\theta_q$ indicates that circling the tip along a path enclosing the defect is equivalent to rotate $\theta_q$ along a closed Fermi surface. Therefore, the phase shift accumulated over a closed scanning path enclosing the defect in monolayer graphene equals to $2\pi$.

Figures 5(d) and 5(e) show atomically resolved STM images of an individual atomic-scale defect at the A and B sublattices of monolayer graphene, respectively. Both the STM images exhibit obvious triangular $\sqrt{3} \times \sqrt{3} R\, 30°$ interference patterns, which are attributed to the charge density modulation induced by the quantum interference of the incident and reflected electronic waves. From the fast Fourier transform (FFT)-filtered STM images, there are robust $|N| = 2$ additional wavefronts in the modulated charge densities for a selected direction of the intervalley scattering, and the directions of the additional wavefronts for defects located at the A and B sublattices are different, as shown in Figs. 5(f) and 5(g). Such results obviously demonstrate that the atomic-scale defect at the A and B sublattices of monolayer graphene can be regarded as pseudospin-mediated atomic-scale vortices with the angular momentum of $l = +2$ and $l = -2$, respectively.[41]

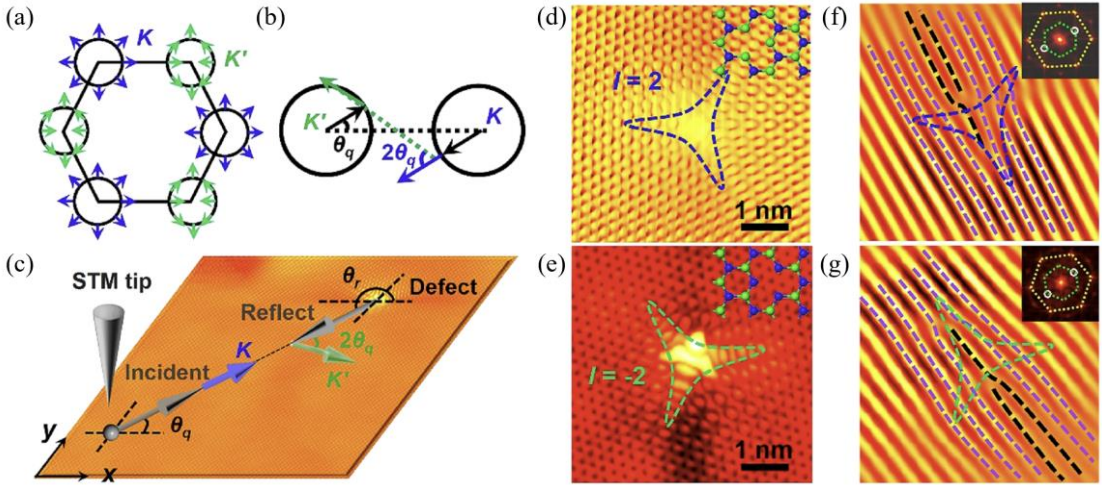

**Fig. 5.** Intervalley scattering induced by an individual atomic-scale defect in monolayer graphene. (a) Pseudospin textures along the Fermi surfaces in monolayer graphene. (b) Intervalley scattering process in graphene. (c) The real space representation of the intervalley scattering in monolayer graphene. The blue and green arrows indicate the direction of pseudospin in K′ and K valley, respectively. (d,e) STM images of an individual atomic-scale defect in monolayer graphene, with the defect located at the A and B sublattices, respectively. (f,g) FFT-filtered STM images along the direction of intervalley scattering indicated by white circles. Insets: FFT of the STM images. The outer hexangular spots and inner bright spots correspond to the reciprocal lattice of graphene and the interference of the intervalley scattering, respectively. Reproduced with permission from Ref. [40]. Copyright 2021, American Chemical Society.

Figure 6 shows the quantum interferences of the multiple atomic-scale vortices with the angular momentum $l = 2$ and/or $l = -2$, which are generated by the atomic-scale defects located at the A and/or B sublattices of monolayer graphene.[40] Although there is an obvious difference in the additional wavefronts around each defect as changing the distance between the defects, the total additional wavefronts are robust, regardless of their positions. In the $|A| = |B|$ case, where $|A|/|B|$ is the number of vortices generated by the defects at the A/B sublattices of monolayer graphene, the interfering vortices cancel each other, resulting in zero total angular momentum $|l| = 0$ and $|N| = 0$ additional wavefronts. While in the $|A| \neq |B|$ case, the interfering vortices show aggregate chirality and angular momenta similar to a single vortex of the majority with $|l| = 2$, thus generating $|N| = 2$ additional wavefronts. All the results can be well understood with the picture of vortices.[40]

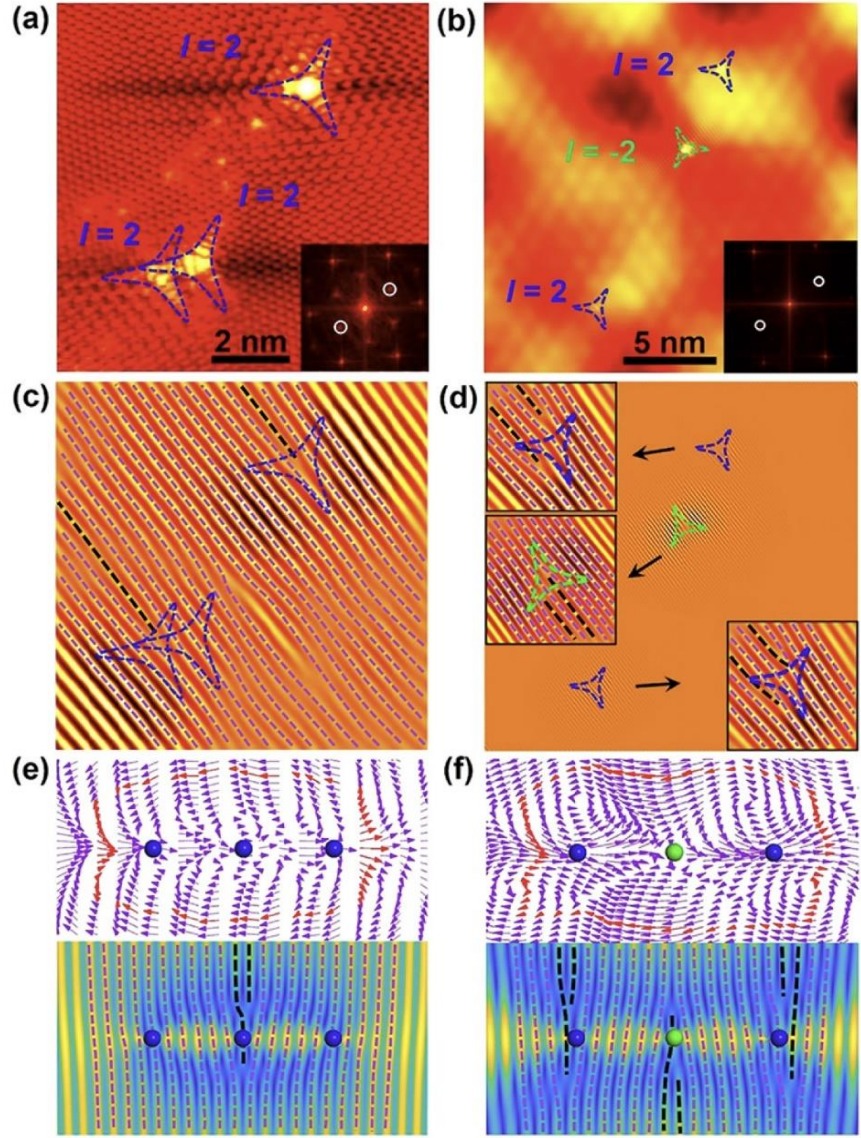

**Fig. 6.** Intervalley scattering induced by three atomic-scale defects in monolayer graphene. (a,b) Typical STM images of A-A-A and A-B-A defects in monolayer graphene, respectively. (c,d) FFT-filtered STM images along the marked direction of the intervalley scattering. (e,f) Upper panels: the structures of the interacting vortices. Bottom panels: the interference patterns between three vortices and a plane wave propagating rightward. Reproduced with permission from Ref. [40]. Copyright 2021, American Chemical Society.

The Berry phase signatures obtained from the intervalley scattering patterns induced by an atomic-scale defect in bilayer graphene are quite different from that in monolayer graphene. In Bernal-stacked bilayer graphene, the pseudospin rotates by the angle of 4π along a closed Fermi surface, yielding the winding number $W = 2$ and the Berry phase $\gamma = W\pi = 2\pi$. In such a case, an atomic-scale defect in bilayer graphene can be regarded as a phase singularity that generates a pseudospin-mediated atomic-scale vortex with the angular momentum $|l| = 4$, and therefore, an atomic-scale defect in Bernal-stacked bilayer graphene is expected to result in $|N| = 4$ additional wavefronts in the modulated charge densities during the intervalley scattering process. However, the atomic-scale defect in Bernal-stacked bilayer graphene is expected to redistribute the numbers of additional wavefronts in the two graphene layers, and such redistributions are strongly dependent on the located sublattice site of the defect. As a result, there are $|N| = 4$, 2, and 0 additional wavefronts in the topmost graphene layer, as shown in Fig. 7.

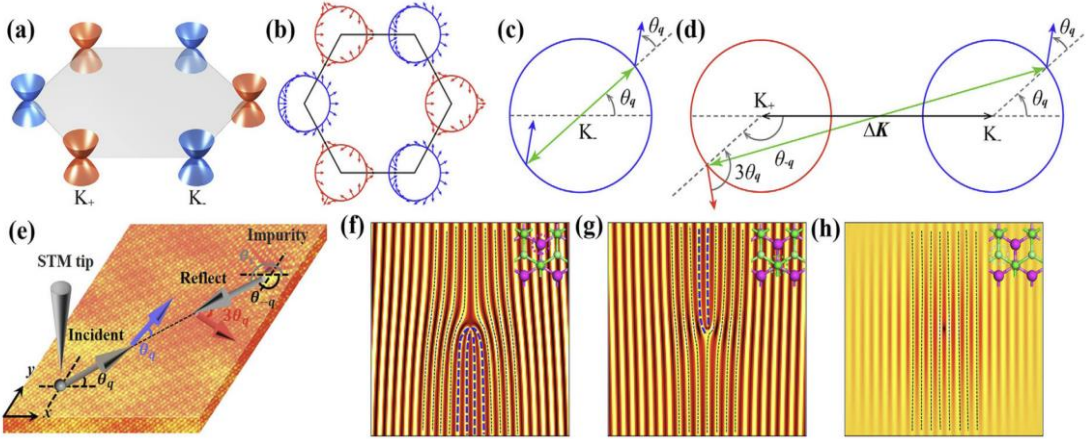

**Fig. 7.** Intervalley scattering induced by an individual atomic-scale defect in bilayer graphene. (a) Low-energy band structures around the Brillouin zone corners of bilayer graphene. (b) Pseudospin textures along the Fermi surfaces of bilayer graphene. (c,d) Schematic intravalley and intervalley scattering processes in bilayer graphene. (e) The real space representation of the intervalley scattering process. The red and blue arrows denote the pseudospins of reflected and incident quasiparticles. (f-h) Charge density oscillations of the topmost graphene layer induced by the intervalley scattering. Reproduced with permission from Ref. [39]. Copyright 2021, American Physical Society.

These results can also be extended to the multilayer graphene systems. For example, in the ABC-stacked trilayer graphene, the $W = 3$ winding number and $\gamma = 3\pi$ Berry phase results in $|N| = 6, 4, 2$, and $0$ additional wavefronts in the topmost graphene layer when the atomic-scale defects are located at different sublattice sites. More generally, the $\gamma = l\pi$ Berry phase of rhombohedral $l$-layer graphene is expected to generate $|N| = 2l, 2l - 2, \ldots, 0$ additional wavefronts for the atomic-scale defect located at different sublattices.[39] Therefore, these works provide a comprehensive measurement of the Berry phase in graphene systems, and can also be applied to other two-dimensional materials.

## 4. Local spin and/or valley polarized states in graphene

By applying an external magnetic field perpendicular to the surface of monolayer graphene, the massless Dirac fermions are expected to condense into a series of Landau levels with the Landau level energy $E_N = sgn(N)\sqrt{2e\hbar v_F^2[N]B} + E_0$. Here $E_0$ is the energy of the Dirac point, $B$ is the magnetic field, $N$ is the Landau level index, $e$ is the electron charge, $\hbar$ is reduced Planck's constant, and $v_F \approx 1 \times 10^6 \, m/s$ is the Fermi velocity.[42-47] In pristine monolayer graphene, each Landau level hosts fourfold degeneracies due to the spin and valley degrees of freedom. If the valley degrees of freedom in monolayer graphene are removed, the $N = 0$ Landau level will split into two valley-polarized peaks. While if both the valley and spin degrees of freedom are removed, the $N = 0$ Landau level will split into four peaks with fully spin-valley polarization (Fig. 8). Therefore, the Landau levels in graphene can be utilized to directly detect the subtle symmetry broken states induced by an individual atomic-scale defect with high spatial resolution.

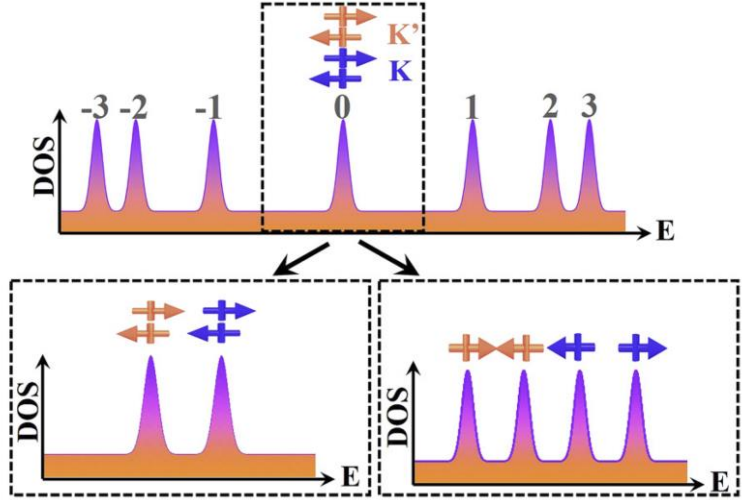

**Fig. 8.** DOS of Landau levels in pristine monolayer graphene under perpendicular magnetic fields. The $N = 0$ LL exhibits fourfold valley and spin degeneracies. If the valley degrees of freedom in monolayer graphene are removed, the $N = 0$ Landau level will split into two valley-polarized peaks. If both the valley and spin degrees of freedom are removed, the $N = 0$ Landau level will split into four peaks with fully spin-valley polarization. Reproduced with permission from Ref. [8]. Copyright 2020, American Physical Society.

Figure 9(a) shows atomically resolved STM image and the corresponding STS spectra recorded around an individual nitrogen-atom dopant in graphene. There is a triangle-like topographic feature in the STM image, with the maximal out-of-plane height of about 85 pm around the nitrogen-atom dopant.[6,48] In the absence of external magnetic fields, the STS spectrum recorded at the site of the nitrogen-atom dopant exhibits a resonant peak at about 0.5 eV above the Fermi energy, together with a prominent electron-hole asymmetry.[6,48] Once applying a $B = 8$ T magnetic field perpendicular to the graphene surface, the STS spectra recorded away from the nitrogen-atom dopant exhibit a well-defined Landau quantization of the massless Dirac fermions. As approaching the nitrogen-atom dopant, the local sublattice symmetry of graphene is broken, and hence, $N = 0$ Landau level gradually splits into two valley-polarized peaks, with their energy separations reaching the maximum of about 20 meV when recorded at the site of the nitrogen-atom dopant (Fig. 9(b)).[49] The magnetic fields further enlarge the energy separations of such valley splitting around the nitrogen-atom dopant, yielding a linear relation with the slope of 2.5 ± 0.3 meV/T.

In contrast, an individual hydrogen-atom absorbed on graphene can induce the carbon atom at the adsorption site distorting from *sp²* to *sp³* hybridization, thus resulting in an obvious protrusion with the lateral height up to about 200 pm in the STM image, as shown in Fig. 9(c). Such an out-of-plane atomic configuration around the hydrogen-atom absorption, on the one hand, locally breaks the sublattice symmetry of graphene and generates a valley-polarized state, on the other hand, significantly promotes the localized interaction between π and σ bands, resulting in the significant enhancement of atomic SOC around the adsorption at the nanoscale.[7,50,51] Figure 9(d) shows the representative STS spectra recorded in the vicinity of the hydrogen-atom absorption under $B = 8$ T magnetic field perpendicular to the graphene surface. As we can see, the $N = 0$ Landau level splits into four peaks, indicating that both valley and spin degeneracies are lifted. Moreover, the nonzero Landau levels split into two peaks, exhibiting a complex relation as a function of $B$, with the energy evolution satisfying

$E_{K,\downarrow} = \Delta + \lambda_A, \ E_{K,\uparrow} = \Delta - \lambda_A,$

$E_{K',\downarrow} = -\Delta + \lambda_B, \ E_{K',\uparrow} = -\Delta - \lambda_B.$

Here $\Delta$ is the potential difference between the nearest carbon site induced by the hydrogen-atom absorption, ↑ and ↓ donate the spin-up and spin-down electrons, $\lambda_A$ and $\lambda_B$ are the strength of atomic SOC on the hydrogen-atom absorbed carbon site and the nearest carbon sites, respectively. The measured atomic SOC around the hydrogen-atom absorbed on graphene is about 5 meV, several orders of magnitude larger than that in pristine graphene.[52,53] In addition, the spatial extension of the enhanced SOC is about 1 nm, much smaller than that of the sublattice symmetry breaking of about 2.5 nm. In addition, the transport measurements demonstrated that the atomic SOC can be greatly enhanced in weakly hydrogenated graphene systems, yielding the average atomic SOC of about 2.5 meV for the case of 0.05% hydrogen-atom absorptions. These results highlight the way to tailor various broken-symmetry states such as SOC in graphene at the nanoscale.[54-56]

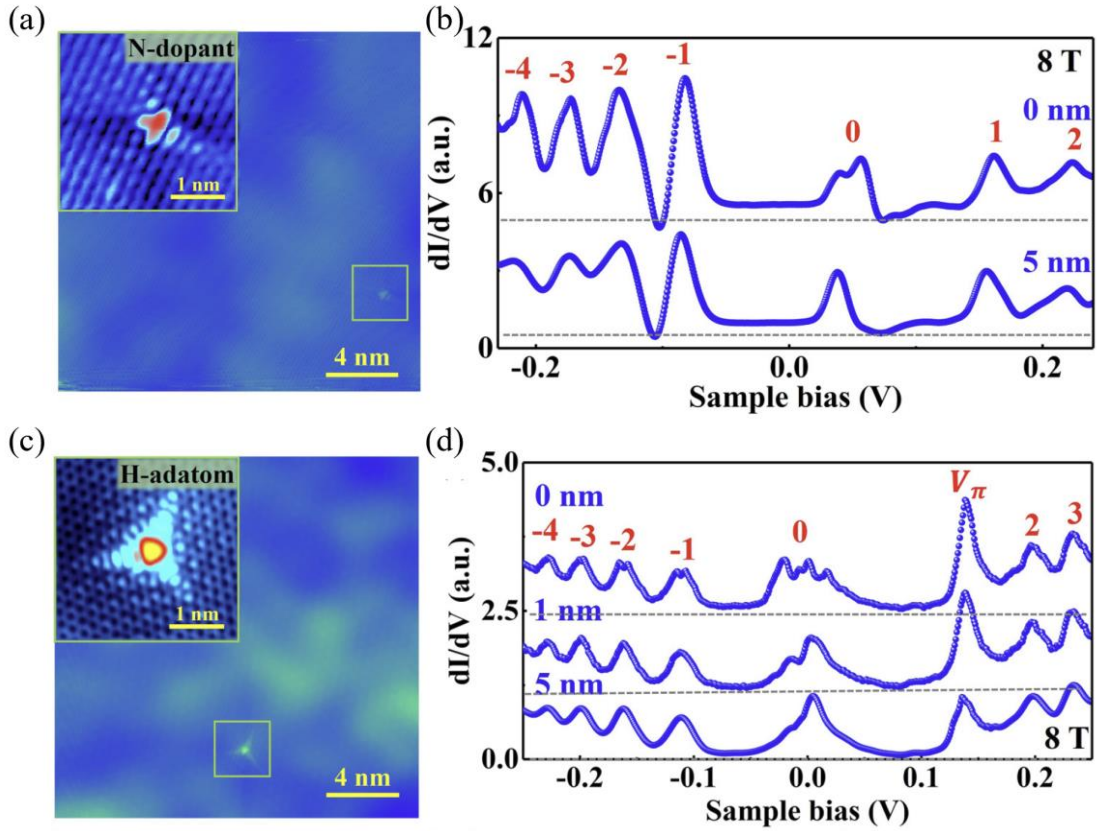

**Fig. 9.** Spin and/or valley polarized states induced by an individual atomic-scale defect in graphene. (a) STM topography of monolayer graphene with an individual nitrogen-atom dopant. (b) Spatially resolved STS spectra recorded around the nitrogen-atom dopant. (c) STM topography of monolayer graphene with an individual hydrogen-atom absorption. (d) Spatially resolved STS spectra recorded around the hydrogen-atom absorption. Reproduced with permission from Ref. [8]. Copyright 2020, American Physical Society.

## 5. Conclusions and perspectives

In summary, several kinds of atomic-scale defects in graphene, including single-carbon vacancy, nitrogen-atom dopant, and hydrogen-atom chemisorption, can efficiently lift the local spin and/or valley degeneracies, resulting in exotic broken symmetry states. All these types of the atomic-scale defects locally break the sublattice symmetry of graphene, and therefore, generate intervalley scattering process and valley polarized states in graphene. Typically, an individual hydrogen-atom chemisorbed on graphene further leads to an enhanced atomic SOC, thus resulting in a fully spin and valley polarized states. Moreover, an individual hydrogen-atom chemisorbed on graphene and an isolated single-carbon vacancy in graphene can generate local magnetic moments, and such magnetic moments can be controllably manipulated by tuning the gate voltage and the local curvature.

Expectatively, there still exist many fundamental mysteries hidden in the defect-induced broken symmetry states in graphene. One of the key challenges is to find out what are the ground states of graphene with different types of atomic-scale defects and their response to the external magnetic fields or gate voltages. Corresponding theoretical calculations need to be carried out to capture the intrinsic physics and predict new broken symmetry states induced by atomic-scale defects in graphene and other two-dimensional materials. Moreover, there is an urgent need for controllable introduction of a certain type of atomic-scale defects with desired densities and patterns by means of optimized growth parameters during the CVD process and the STM-tip manipulation techniques.[59] On the one hand, the local magnetic moments induced by atomic-scale defects in graphene can ferromagnetically or antiferromagnetically couple with one another, depending on the sublattice sites of the vacancies, which are expected to realize global giant magnetic moments. On the other hand, significant enhancement of global SOC and spin/valley polarization induced by atomic-scale defects arranged in array are expected to realize spin and valley Hall effects, thus paving the way for further application in spintronics and valleytronics.

**Acknowledgments**

This work is financial supported by National Natural Science Foundation of China (Nos. 92163206 and 61725107), National Key Research and Development Program of China (2020YFA0308800), Beijing Natural Science Foundation (No. Z190006), and China Postdoctoral Science Foundation (No. 2021M700407).

**Competing financial interests**

The authors declare no competing interests.